\documentclass[12pt]{article}

\usepackage{amsmath}
\usepackage{amsfonts}
\usepackage{amssymb}
\usepackage{amsthm}
\usepackage{setspace}
\usepackage{color}
\usepackage{subcaption}
\usepackage{caption}
\usepackage{enumerate}
\usepackage[hidelinks]{hyperref}
\usepackage{graphicx}
\usepackage[hidelinks]{hyperref} 
\usepackage[nottoc]{tocbibind}
\usepackage[utf8]{inputenc} 

\newcommand\ee{\end{equation}}
\newcommand\be{\begin{equation}}
\newcommand\eea{\end{eqnarray}}
\newcommand\bea{\begin{eqnarray}}

\newcommand\mpl{M_{\rm pl}}
\newcommand\comment[1]{}
\newcommand\expect[1]{\left\langle #1 \right\rangle}
\newcommand\bsb{\boldsymbol}

\comment{
\hypersetup{
colorlinks=true,
citecolor=DarkBlue,
linkcolor=DarkBlue,
urlcolor=DarkBlue,
}}

\def\O{\mathcal{O}}

\def\d{\partial}

\def\E{{\mathcal{E}}}

\def\vep{\varepsilon}

\def\p{{\bsb p}}

\def\k{{\bsb k}}
\def\x{\bsb x}

\def\ep{\epsilon}

\def\P{\bsb P}

\begin{document}

\begin{center}

  {\Large\bf Loosely coupled particles in warm inflation}

\vskip 1 cm
{Mehrdad Mirbabayi}
\vskip 0.5 cm

{\em International Centre for Theoretical Physics, Trieste, Italy}

\vskip 1cm

\end{center}
\noindent {\bf Abstract:} {\small Cosmological perturbations in warm inflation are conventionally described by an EFT. It consists of the inflaton field coupled to a radiation fluid and applies to energy scales of order Hubble $H$ and inflaton decay width $\gamma$, but well below temperature $T$. This EFT lacks a proper treatment of inflatons produced with energy $\sim T$. While the direct contribution of these ``UV inflatons'' to the super-horizon curvature perturbations is subdominant because of the cosmological redshift, for $\gamma \sim H$, their indirect contribution is potentially important---by their effect on the equation of state, by being exchanged as long-lived intermediate states, and by sourcing gravitational waves with their sizable anisotropic stress. To include these effects, we add to the EFT the Boltzmann distribution of the gas of UV inflatons and compute some of the corrections.

\vskip 1 cm

\section{Introduction}

Warm inflation is a model in which interactions between the inflaton field $\phi$ and a thermal bath slow down the inflaton and fuel the bath against the Hubble expansion \cite{Moss,Maeda,Fang}. To our knowledge, the only successful realization of this idea is via sphaleron heating, when the inflaton is an axion and the thermal bath is a pure Yang-Mills (YM) plasma \cite{Berghaus}. 

However, regardless of the underlying mechanism, warm inflation is known to have distinct observational predictions that are dictated by a relatively simple EFT consisting of the inflaton field coupled to a radiation fluid \cite{Bastero-Gil,wng}. For instance, for a given potential $V(\phi)$ warm inflation predicts different scalar-tilt $n_s$ and smaller tensor-to-scalar ratio $r$. It also predicts an irreducible and potentially observable level of non-Gaussianity with $f_{\rm NL}\sim 1-50$. And the leading effects are all controlled by a single model parameter: the ratio of the thermal friction to Hubble: $\gamma/H$. 

The goal of this paper is to point out a so far missing component of the above-mentioned EFT, which can lead to appreciable corrections when $\gamma\sim H$. So we begin by describing in more detail what has been included so far. We have an inflaton field $\phi$ and a fluid described by its energy momentum tensor
\be
T^{\rm rad}_{\mu\nu} = \frac{4}{3}\rho u_\mu u_\nu + \frac{1}{3}\rho g_{\mu\nu}+\cdots
\ee
where dots include dissipation and corrections to the equation of state. They are coupled together, and to gravity, via
\be\label{coveq0}\begin{split}
D^2 \phi - V'(\phi)& = \gamma u^\mu \d_\mu \phi + \xi,\\[10pt]
D^\mu T^{\rm rad}_{\mu\nu}& =- \d_\nu \phi (\gamma u^\mu \d_\mu \phi +\xi),\\[10 pt]
G_{\mu\nu} &= 8\pi G T^{\rm tot}_{\mu\nu},
\end{split}
\ee
where $D_\mu$ is the covariant derivative, $\gamma$ is a function of the local temperature (or, equivalently, of the fluid density $\rho$):
\be
\gamma = \gamma_0 \left(\frac{\rho}{\rho_0}\right)^{3/4}.
\ee
$\xi$ is a noise whose 2-point function in the infrared is fixed by the fluctuation-dissipation theorem. That is, within the EFT
\be\label{xi2pf}
\expect{\xi(x_1)\xi(x_2)} \approx 2\gamma T \delta^4(x_2 - x_1),\qquad \text{(physical coordinates)}.
\ee
This EFT, like any other, relies on a separation of the scale of interest $H$ from the UV scale $T$, and it has free parameters to incorporate the unspecified details of the UV physics. For instance, the deviation of the equation of state $p/\rho$ from $1/3$, the viscosity coefficients, the higher point statistics of $\xi$, etc. These are by and large irrelevant for the long-wavelength (i.e. cosmological) predictions of the model. 

However, there is a subtlety that distinguishes this model from a typical EFT. Given the background expansion, the UV fluctuations of the inflaton field are neither in Bunch-Davies vacuum (unless $\gamma$, which is a measure of the interaction rate between the inflaton and the bath, is $\ll H$) nor in a thermal state (unless $\gamma \gg H$). When $\gamma \sim H$, these degrees of freedom can neither be integrated out in the Wilsonian sense nor in the hydrodynamic sense. 

Aren't these automatically included in the model \eqref{coveq0}? After all, we are talking about the UV fluctuations of $\phi$ that is already part of the system. Furthermore, EFTs usually tell us if they are missing something. 

In fact, the noise term, with a local correlator as in \eqref{xi2pf}, does lead to the production of inflatons at all scales, which is a potential concern because it is not supposed to be accurate in the UV. In \cite{wng}, we discarded this problem by arguing that it is the infrared part of the spectrum that dominates the answer for the tree-level cosmological correlators. However, an attempt to include the loop corrections reveals the peculiarity of our system. In particular, one encounters loops in which on-shell UV excitations of $\phi$ propagate. Unless $\gamma \gg H$, there is a large separation between their production and absorption and hence these UV contributions cannot be absorbed in the local counter-terms. We cannot hope our effective hydro description to give an accurate description at energy $\sim T$, and its lack of freedom to absorb the UV contributions is a clear shortcoming. 

To get a feeling about the importance of the gas of UV inflatons, it is useful to estimate their energy density, which for $\gamma < H$ is
\be
\rho_{g} \sim \frac{\gamma}{H} T^4.
\ee
When $\gamma \ll H$, this is a small contribution compared to $\rho_{\rm rad}$ and the corresponding effect on correlation functions is expected to be similarly suppressed by powers of $\gamma/H$. When $\gamma \gg H$, UV inflatons thermalize with the bath and become part of the radiation fluid. We expect a nontrivial contribution when $\gamma \sim H$. In the context of axion inflation coupled to $SU(N)_{\rm YM}$, this corresponds to novel effects suppressed by the number of Yang-Mills degrees of freedom $\frac{1}{2(N^2-1)}$, i.e. never bigger than about $10\%$.

In the rest of this paper, we formulate our proposed resolution in section \ref{sec:gas}. In section \ref{sec:pert}, we set up the perturbation theory and in particular write the equations for linear scalar and tensor perturbations. Warm inflation generically predicts negligible tensor-to-scalar ratio $r$ because it greatly enhances the scalar power. When $\gamma \sim H$, the gas of almost free-streaming inflatons sources gravitational waves more efficiently than the tightly coupled radiation fluid. This is perhaps the optimum regime for enhancing $r$ in warm inflation, as expected from general arguments in \cite{GW,Qiu}. In section \ref{sec:ph4}, we study the explicit example of $\phi^4$ warm inflation with $\gamma \simeq 5 H$. We find tensor modes to be still dominated by vacuum fluctuations and negligibly small. We conclude in section \ref{sec:con}.

\section{A gas of inflatons}\label{sec:gas}
The original fluid-inflaton EFT for warm inflation is incomplete because it cannot give an accurate description of inflaton production at the temperature scale, and this process has an important effect on cosmological observables when $\gamma \sim H$. This problem can be fixed as follows. We add to the inflaton-fluid EFT a gas of inflaton particles, with absorption and production rates as free UV inputs. The full system of equations now includes the Boltzamann equation for the distribution function of the inflaton gas. This is a degenerate description of the inflaton field, but it introduces enough freedom to allow matching with the underlying microscopic model. 

It is uncommon to use phase-space distributions and the Boltzmann equation in the context of inflation because one often deals with condensates. However, it is the standard formalism to study, among others, the thermal freeze-out of QCD axions during the cosmic history \cite{Villadoro}. In such cases, particle wavelengths are much shorter than the mean-free-length. Production of inflatons by the YM plasma in minimal warm inflation is a close analog of axion production by the QCD plasma, and most of this section is a review of the well-known results. In particular, the spatial average of the inflaton distribution satisfies
\be\label{Blz}
\d_t f_0(t,\p) =\Gamma^<(E_p) (1+f_0(t,\p))- \Gamma^>(E_p) f_0(t,\p),
\ee
where up to corrections that are suppressed by both slow-roll parameters and $H/T$, inflaton can be approximated as being massless, with energy $E_p = p^0= |\p|/a$. $\Gamma^<$ and $\Gamma^>$ are, respectively, the production and absorption rates per unit phase space of inflatons by the thermal bath. Implicitly, they also depend on $T$, and satisfy
\be\label{nonpert}
\Gamma^<(E)  = e^{-E/T} \Gamma^>(E),
\ee
using which \eqref{Blz} can be written as
\be
\d_t f_0(t,\p) = \bar \Gamma(E_p) [f_{\rm eq}(\p)-f_0(t,\p)],
\ee
where $f_{\rm eq}(\p) = 1/(e^{E_p/T}-1)$, and
\be
\bar \Gamma(E) = \Gamma^>(E) \left(1- e^{-E/T}\right).
\ee
In sphaleron-based warm inflation, we know that $\alpha_{\rm YM}$ cannot be extremely small if the observed scalar power is to be matched \cite{wng}. For moderately small $\alpha_{\rm YM}$, there is no big hierarchy between $\bar\Gamma(E)$ at different energies, hence
\be\label{Gg}
\bar\Gamma(E) \sim \bar\Gamma(0)=\gamma\propto T^3.
\ee
Inflaton particles are created and absorbed by the bath, hence the background equation for the fluid energy is modified to 
\be
\dot \rho_0 + 4 H \rho_0= \gamma \dot\phi_0^2 - \int \frac{d^3\p}{(2\pi a)^3} E_p \Gamma^>(E_p)[e^{-E_p/T}(1+f_0)-f_0],
\ee
and the Friedmann equation reads
\be
H^2 = \frac{8\pi G}{3} \left[\frac{1}{2}\dot\phi_0^2 + V(\phi_0)+ \rho_0 + \rho_g\right],
\ee
where $\rho_g$ is the energy density of the inflaton gas
\be\label{rhog}
\rho_g = \int \frac{d^3\p}{(2\pi a)^3} E_p f_0.
\ee
The background equation for the inflaton remains the same as before:
\be\label{bg}
\ddot{\phi_0}+(3 H +\gamma) \dot{\phi_0} + V'(\phi_0) = 0.
\ee
Using the approximation \eqref{Gg} one can numerically check that these equations have a warm slow-roll attractor solution when $V(\phi)$ is sufficiently flat. An example is discussed in section \ref{sec:ph4}. On this attractor the distribution written as a function of physical rather than comoving momenta, namely
\be\label{F0}
F_0(t,\P) \equiv f_0(t,a\P)
\ee
is time-independent, up to slow-roll corrections. It satisfies
\be\label{feq}
-H \P\cdot \nabla_P F_0 \approx \bar\Gamma(|\P|) [F_{\rm eq}- F_0].
\ee
When $\bar\Gamma \gg H$, the expression in the square brackets on the RHS is small, and therefore the inflaton gas reaches thermal equilibrium with the plasma, $F_0(t,\P) \approx \frac{1}{e^{|\P|/T}-1}$. Thus our addition is equivalent to changing the effective number of degrees of freedom in the bath. Away from this limit, its effect even on the background solution is not exactly degenerate with radiation. 

\section{Cosmological perturbations}\label{sec:pert}
To study perturbations, we introduce noise terms $\xi,\sigma^<,\sigma^>$ that keep track of individual production and absorption events, and (ignoring metric perturbations) obtain the following system
\be \label{coveq}\begin{split}
D^2 \phi - V'(\phi)& = \gamma u^\mu \d_\mu \phi + \xi,\\[10pt]
D^\mu T^{\rm rad}_{\mu\nu}& =- \d_\nu \phi (\gamma u^\mu \d_\mu \phi +\xi) + X_\nu,\\[10pt]
\frac{1}{E_p} p^\mu\d_\mu f(t,\x,\p) &= \sigma^<(t,\x,\p)-\sigma^>(t,\x,\p),
\end{split}
\ee
where $p^\mu=E_p(1,\hat p/a)$, $\xi$ is the noise that couples the {\em inflaton field} to the bath and satisfies \eqref{xi2pf}, and given that inflaton particles are emitted and absorbed by the bath, we have added
\be
X_\mu \equiv \int \frac{d^3 \p}{(2\pi a)^3} \ p_\mu (\sigma^<(t,\x,\p) - \sigma^>(t,\x,\p)),
\ee
to the RHS of the fluid energy-momentum evolution equation. Inflaton emission and absorption events have a characteristic size of $\O(1/T)$. We always work in the regime $\gamma\ll T$, and therefore can focus on time-scales $\gg 1/T$ but shorter than $1/\gamma$ to define $\sigma$ as having Poisson statistics:
\be\label{eqtime}\begin{split}
  \expect{\sigma^<(t_1,\x_1,\p_1)\cdots \sigma^<(t_n,\x_n,\p_n)}&= \frac{(-u\cdot p_{1})\Gamma^<(-u\cdot p_1)}{E_1}
  (1+f(t_1,\x_1,\p_1))
  \\[10pt]
  &~~~~~~\prod_{i=2}^n\delta(t_i-t_1)
(2\pi)^3\delta^3(\x_i-\x_1)\delta^3(\p_i-\p_1),\\[10pt]
  \expect{\sigma^>(t_1,\x_1,\p_1)\cdots \sigma^>(t_n,\x_n,\p_n)}&= \frac{(-u\cdot p_{1})\Gamma^>(-u\cdot p_1)}{E_1} f(t_1,\x_1,\p_1)\\[10pt]
  &~~~~~~~\prod_{i=2}^n\delta(t_i-t_1)
(2\pi)^3\delta^3(\x_i-\x_1)\delta^3(\p_i-\p_1),
\end{split}
\ee
where $-u\cdot p_1=- u_\mu p_1^\mu$ is the energy measured in the rest-frame of the fluid, and $E_1 = |\p_1|/a$ is the energy in the cosmological background frame. Since the phase-space density is a Lorentz scalar, the last line of \eqref{coveq} implies that $E_p\sigma(t,\x,\p)$ is a scalar too. This is consistent with the above correlations, given that $\Gamma^{\gtrless}$ are defined in the rest-frame of the fluid, and hence are Lorentz scalars, and $a(t)^3 E_1\delta^3(\p_1-\p_i)$ and $\delta(t_1-t_i) \delta^3(\x_1-\x_i)/a(t)^3$ are both Lorentz invariant. 

The assumption that there are uncorrelated single particle production and absorption events is motivated by the minimal warm inflation. In that model, the Peccei-Quinn scale is much bigger than $T$, leading to a suppression of processes with more incoming and outgoing inflatons. 

The system of equations \eqref{coveq}, with the prescribed statistics for $\xi$ and $\sigma^{\gtrless}$ is a well-formulated stochastic system, as was the original inflaton-fluid system. It is significantly harder to evolve because it involves a phase-space distribution, but in principle, it can be solved to find the cosmological correlation functions.
\subsection{Linear perturbations; scalar modes}
To compute the scalar power, we linearize the above system of equations. We still have translation symmetry, and hence it is convenient to work in momentum space ($\k$ dual to the comoving coordinates $\x$). We define the perturbations via
\be\label{def}
\phi = \phi_0 + \dot\phi_0 \Phi,\quad
T^{~0}_{{\rm rad}~0} = -\rho_0 (1+ \E),\quad
\d_i T^{~i}_{{\rm rad}~0} =  - \frac{4\rho_0}{3a^2} \nabla^2 \Psi.
\ee
In addition, there is also $F_\k(t,\P)=f_\k(t,a\P)$, where $\P$ is the physical momentum of inflaton particles. Neglecting slow-roll corrections, superhorizon curvature perturbations are related to $\Phi$ via $\zeta_\k = - H\Phi_\k$.

Next, we decompose 
\be
\sigma^\gtrless = \expect{\sigma^{\gtrless}}_{\Phi,\E,\Psi,F} + \hat \sigma^{\gtrless},
\ee
and neglect slow-roll corrections to find the following equations for linear perturbations
\be\label{linEq}\begin{split}
\d_t F_\k - H\P\cdot\nabla_P F_\k+ i\frac{\hat p \cdot \k}{a}F_\k
&=\frac{1}{4} \E_\k\left[(T\d_T\bar\Gamma-x\d_x \bar\Gamma)(F_{\rm eq}-F_0)-\bar\Gamma x\d_xF_{\rm eq}\right]
\\[10pt]
&~ - i \frac{\hat p\cdot \k}{a}\Psi_\k\left[(\bar\Gamma+x\d_x \bar\Gamma)(F_{\rm eq}-F_0)
+\bar\Gamma x\d_x F_{\rm eq}\right]
\\[10pt]
&~ -\bar\Gamma F_\k + \hat\sigma^<_\k - \hat \sigma^>_\k\\[10pt]
\dot \E_\k +(1-3 y) H  \E_\k - \frac{4 k^2}{3a^2 } \Psi_\k - 8 (1+y) H \dot\Phi
& = \frac{4 H }{\gamma \dot\phi_0} (1+y) \xi
-\int \frac{d^3\P}{(2\pi)^3\rho_0} P(\sigma^<_\k-\sigma^>_\k)\\[10pt]
\dot \Psi_\k+ 3H \Psi_\k+ \frac{1}{4} \E+ 3(1+y) H\Phi
&=\frac{3i a}{4k^2}\int \frac{d^3\P}{(2\pi)^3\rho_0} (\k \cdot \P) (\sigma^<_\k-\sigma^>_\k)\\[10pt]
\ddot \Phi + (3H + \gamma) \dot\Phi+ \frac{3}{4} \gamma \E + \frac{k^2}{a^2}\Phi
& =- \frac{1}{\dot\phi_0} \xi,
\end{split}
\ee
where $P=|\P|$, $x=P/T$ and in the second and third equations $(\sigma^<_\k - \sigma^>_\k)$ can be replaced by the RHS of the first equation. We also defined the ratio of the average gas energy density \eqref{rhog} to the bath density:
\be
y = \frac{\rho_g}{\rho_0},
\ee
in terms of which,
\be
\gamma \dot\phi_0^2 \approx 4 H \rho_0(1+y)
\ee
up to slow-roll corrections. Given $\bar\Gamma/H$ the slow-roll equilibrium distribution $F_0$ is uniquely fixed by \eqref{feq} as a function of $P/T$. Therefore, the ratio $y$ is fixed once we know the number of degrees of freedom in the plasma $g_*$ [$= 2(N^2-1)$ for pure $SU(N)_{\rm YM}$], in terms of which $\rho_0 = \frac{\pi^2}{30} g_* T^4$.

\subsection{Tensor modes}
Tensor modes $\chi_{ij}$ are defined by writing $g_{ij}= a^2 (\delta_{ij}(1+2\zeta) + \chi_{ij})$ with $\chi_{ii} = 0,\d_i \chi_{ij}=0$. In momentum space, we expand
\be
\chi_{ij} = \vep_{ij}^{++} \chi_+ + \vep^{--}_{ij} \chi_-,
\ee
where for $\hat k = \hat z$, $\vep^{++}_{ij} = \frac{1}{2} \vep^+_i \vep^+_j$, ${\vep^+} = \hat x + i \hat y$ and $\vep^{--} = (\vep^{++})^*$. At linear order, $\chi _+$  satisfies
\be\label{chi}
\ddot \chi_+ +3 H \dot\chi_{+} +\frac{k^2}{a^2}\chi_{+}= 8\pi G a^2 \vep_{ij}^{--} T^{ij}.
\ee
When $\gamma\sim H$, the anisotropic stress of the inflaton gas is correlated at cosmological scales and hence efficiently sources the primordial gravitational waves. In terms of the phase space distribution
\be\label{T--}
a^2 \vep_{ij}^{--} T^{ij} = -\frac{1}{5 \sqrt{6}\pi^2} \int_0^\infty dP P^3 F_{\k,2,2},
\ee
where we expanded $F_\k$ in terms of spherical harmonics, 
\be
F_\k(t,\P) = \sum_{l,m} F_{\k,l,m}(t,P) i^l Y_{lm}(\hat P),\qquad
\int d^2\hat P Y_{lm}(\hat P) Y^*_{l'm'}(\hat P) = \frac{4\pi}{2l+1}\delta_{l l'}\delta_{mm'}.
\ee
In the same basis, the 2-point correlation functions of creation/absorption events are
\be\label{siglm}
\expect{\hat \sigma^\gtrless_{\k,l,m}(t,P) \hat \sigma^\gtrless_{\k',l',m'}(t',P')} =N^{\gtrless}
(2l+1)\delta_{ll'}\delta_{m,-m'}\frac{2\pi^2 \delta(P- P')}{a^3 P^2}
\delta(t-t') (2\pi)^3\delta^3(\k + \k') ,
\ee
where at zeroth order in perturbations
\be
N^< =  \frac{\bar \Gamma}{e^{P/T}-1} (1+F_0(P)) ,\qquad N^>  = \frac{\bar \Gamma}{1-e^{-P/T}} F_0(P).
\ee
Naturally, at linear order, scalar and tensor perturbations decouple. In particular, the first equation in \eqref{linEq} reads
\be\begin{split}\label{Flm}
&\d_t F_{\k,l,m}- P \d_P F_{\k,l,m} +\frac{k}{a}
\left[\frac{\sqrt{l^2-m^2}}{2l-1} F_{\k,l-1,m}-\frac{\sqrt{(l+1)^2-m^2}}{2l+3}F_{\k,l+1,m}\right]
\\[10pt]
&=\frac{\delta_{l0}\delta_{m0}}{4} \E_\k\left[(T\d_T\bar\Gamma-x\d_x \bar\Gamma)(F_{\rm eq}-F_0)-\bar\Gamma x\d_xF_{\rm eq}\right]
 -  \frac{k \delta_{l1}\delta_{m0}}{a}\Psi_\k\left[(\bar\Gamma+x\d_x \bar\Gamma)(F_{\rm eq}-F_0)
+\bar\Gamma x\d_x F_{\rm eq}\right]
\\[10pt]
&~ -\bar\Gamma F_{\k,l,m} + \hat\sigma^<_{\k,l,m} - \hat \sigma^>_{\k,l,m},
\end{split}
\ee
which mixes various $l$'s but not various $m$'s. Since the last three equations in \eqref{linEq} only involve $m=0$ and $m=1$, they are decoupled from tensor modes, which are instead sourced by the $m=\pm 2$ components. Equations \eqref{chi} and \eqref{Flm} with the noise spectrum \eqref{siglm} form a well formulated stochastic system that can be solved to find the sourced contribution to the tensor power. 
\section{An example: Warm $\phi^4$ inflation}\label{sec:ph4}
We consider the same example as in \cite{wng}, $V(\phi) =\frac{\lambda}{4} \phi^4$ model, but now include the missing effect of the inflaton gas. Let us define $\hat \phi = \sqrt{\lambda}\phi$, $\hat T = \gamma^{1/3}$ and work in units where $\mpl^2 =\frac{1}{8\pi G} = \frac{1}{\lambda}$. We also assume that $\bar\Gamma = \gamma$. Then the background equations become
\be\begin{split}
\d_t^2 \hat \phi &= - (3 H + {\hat T}^3)\d_t \hat \phi - {\hat \phi}^3,\\[10 pt]
\d_t \hat T & = - H \hat T  + \beta (\d_t \hat \phi)^2 - \beta (\hat\rho_{\rm eq} - \hat\rho_g),\\[10pt]
\d_t F_0 & =  H P\d_P F_0 +{\hat T}^3 (\hat F_{\rm eq}- F_0),\\[10pt]
H^2 &= \frac{1}{3}\left[\frac{1}{2}(\d_t\hat\phi)^2 + \frac{{\hat\phi}^4}{4}+\frac{1}{4\beta}{\hat T}^4+\hat\rho_g\right],\end{split}\ee
where $\beta$ is a constant combination of model parameters 
\be
\beta = \frac{\gamma^{4/3}}{4\lambda \rho_0},
\ee
and $\hat F_{\rm eq} = 1/(e^{P/\hat T}-1)$. We have also defined the normalized density
\be
\hat \rho_g = \frac{15}{4\pi^4 g_* \beta}\int_0^\infty dP P^3 F_0,
\ee
and similarly $\hat \rho_{\rm eq}$ by replacing $F_0\to \hat F_{\rm eq}$, so that at equilibrium $\hat\rho_{g} = \frac{{\hat T}^4}{4g_* \beta}$. With the same choice of $\beta = 10$ as in \cite{wng}, and $g_* = 6$ for $SU(2)_{\rm YM}$, we now obtain at $N_e = 55$ before reheating, defined as $V(\phi)=\rho_0$, 
\be\label{55}
\phi\approx  11.8 \mpl,\qquad \gamma \approx  5.04 H,\qquad  H\approx 40.7 \sqrt{\lambda}\mpl,\qquad \rho_g \approx  0.092 \rho_0.
\ee
to be compared with $\phi\approx 11.6 \mpl,\gamma \approx 5.34 H, H\approx 39.0\sqrt{\lambda}\mpl$ found in \cite{wng}. The fact that $\rho_g<\rho_0/6$ indicates that $\gamma$ is not large enough to bring the gas into thermal equilibrium. 

We numerically solve the perturbation equations (at zeroth order in slow-roll parameters) to find the tensor-to-scalar ratio. This is a computationally demanding task because we have to evolve the gas distribution, hence solve a PDE in $(t,\P)$ variables for every Fourier mode. Furthermore, there are independent noise terms $\hat \sigma^\gtrless$ at every $(t,\P)$, or every $(t,P,l,m)$ in the harmonic basis. The result can be expressed in terms of the Green's functions defined for every independent noise. For instance, $G_{\Phi_\k(t_f)}^{\xi_\k(t)}$ is the resulting $\Phi_\k(t_f)$ for $\xi_\k$ being a delta function at $t$. Choosing $t_f$ such that $k\ll H a(t_f)$, the sourced scalar power reads
\be\label{Pz}\begin{split}
k^3 \expect{\zeta_\k \zeta_{-\k}}_{\xi,\sigma}' &= H^2\int_{-\infty}^{t_f} \frac{dt}{a^3} \left[2\gamma T \left(G_{\Phi_\k(t_f)}^{\xi_\k(t)}\right)^2 
+\sum_{l\geq 0} (2l+1)\int_0^\infty \frac{2\pi^2 dP}{P^2} (N^>+N^<) \left(G_{\Phi_\k(t_f)}^{\hat\sigma_{\k l 0}(t,P)}\right)^2\right]\\[10pt]
&\equiv (A_\xi+ A_\sigma) \frac{H^3}{T^3},\end{split}
\ee
where prime on the correlator indicates that a factor of $(2\pi)^3\delta^3(\k+\k')$ has been dropped. Similarly, for the tensor modes
\be\label{Pchi}\begin{split}
k^3 \expect{\chi^+_\k \chi^+_{-\k}}_\sigma' &= \sum_{l\geq 2} (2l+1) \int_{-\infty}^{t_f} \frac{dt}{a^3} \int_0^\infty \frac{2\pi^2 dP}{P^2} (N^>+N^<) \left(G_{\chi^+_\k(t_f)}^{\hat\sigma_{\k l 2}(t,P)}\right)^2\\[10pt]
&\equiv B_\sigma \frac{H^3}{T^3}.\end{split}
\ee
In our numerics, we include up to $l\leq 6$ with 10 energy bins in the range $T/10<P<10 T$. For the same parameters as in \eqref{55}, and using the background slow-roll solution $\dot\phi^2_0 = 4H(1+y)\rho_0/\gamma$, we find 
\be
A_\xi \approx 50,\qquad A_\sigma \approx 5,\qquad B_\sigma \approx 0.03 \left(\frac{\rho_0}{V}\right)^2,
\ee
resulting in the following sourced contribution to the tensor-to-scalar ratio
\be
r_{\rm sourced} =  \frac{2B_\sigma}{A_\xi+A_\sigma} \approx 4\times 10^{-8}.
\ee
This is about an order of magnitude less than what was obtained in \cite{wng} based on zero-point fluctuations of the tensor modes! 

Warm inflation is way more efficient in exciting scalar modes than tensor modes. Some of this suppression could have been anticipated. The ratio $\rho_0/V$ [$= \frac{\gamma^{4/3}}{\beta \hat\phi^4} \approx 6\times 10^{-3}$ for the values in \eqref{55}] is of order $\ep_H \equiv -\dot H/H^2$. The suppression of $B_\sigma$ by its second power is in agreement with the general argument of \cite{GW}: scalar perturbations are by a factor of $1/\ep_H$ more strongly coupled to a given particle production event. An extra suppression $\propto (\rho_g/\rho_0) \sim 10^{-1}$ is expected because only a fraction of produced particles have large anisotropic stress. 

\section{Conclusions}\label{sec:con}
Warm inflation has universal predictions because of having a large separation of scales that allows an effective description in terms of inflaton field coupled to a fluid. We argued that this EFT is incomplete because it cannot accurately describe inflatons that are produced with energy $\sim T$ and proposed a way to fix this problem by including in the EFT a Boltzmann distribution for these thermally produced inflatons. This formalism can probably be applied whenever loosely coupled particles are produced during inflation. 

Given the absorption rate of inflatons by the thermal bath, $\bar\Gamma(E)$, the new theory is predictive. The new effects are significant when $\bar\Gamma \sim H$ because the inflaton gas has an energy density comparable to that of the thermal bath without being fully thermalized. We have set up a systematic perturbative scheme to compute cosmological perturbations and applied it to calculate scalar and tensor power spectra in a particular example. 

It is conceivable that the new effect alters the shapes of non-Gaussianity found in \cite{wng} and introduce new ones. However, we expect them to be at best $\O(10\%)$ corrections, and unfortunately challenging to compute.

\vspace{0.3cm}
\noindent
\section*{Acknowledgments}

We thank Paolo Creminelli, Andrei Gruzinov and Giovanni Villadoro for useful discussions.

\bibliography{bibwng}

\providecommand{\href}[2]{#2}\begingroup\raggedright\begin{thebibliography}{1}

\bibitem{Moss}
I.~G. Moss, ``{Primordial Inflation With Spontaneous Symmetry Breaking},''
  \href{http://dx.doi.org/10.1016/0370-2693(85)90570-2}{{\em Phys. Lett. B}
  {\bfseries 154} (1985) 120--124}.

\bibitem{Maeda}
J.~Yokoyama and K.-i. Maeda, ``{On the Dynamics of the Power Law Inflation Due
  to an Exponential Potential},''
  \href{http://dx.doi.org/10.1016/0370-2693(88)90880-5}{{\em Phys. Lett. B}
  {\bfseries 207} (1988) 31--35}.

\bibitem{Fang}
A.~Berera and L.-Z. Fang, ``{Thermally induced density perturbations in the
  inflation era},'' \href{http://dx.doi.org/10.1103/PhysRevLett.74.1912}{{\em
  Phys. Rev. Lett.} {\bfseries 74} (1995) 1912--1915},
  \href{http://arxiv.org/abs/astro-ph/9501024}{{\ttfamily
  arXiv:astro-ph/9501024}}.

\bibitem{Berghaus}
K.~V. Berghaus, P.~W. Graham, and D.~E. Kaplan, ``{Minimal Warm Inflation},''
  \href{http://dx.doi.org/10.1088/1475-7516/2020/03/034}{{\em JCAP} {\bfseries
  03} (2020) 034}, \href{http://arxiv.org/abs/1910.07525}{{\ttfamily
  arXiv:1910.07525 [hep-ph]}}.

\bibitem{Bastero-Gil}
M.~Bastero-Gil, A.~Berera, I.~G. Moss, and R.~O. Ramos, ``{Theory of
  non-Gaussianity in warm inflation},''
  \href{http://dx.doi.org/10.1088/1475-7516/2014/12/008}{{\em JCAP} {\bfseries
  12} (2014) 008}, \href{http://arxiv.org/abs/1408.4391}{{\ttfamily
  arXiv:1408.4391 [astro-ph.CO]}}.

\bibitem{wng}
M.~Mirbabayi and A.~Gruzinov, ``{Shapes of non-Gaussianity in warm
  inflation},'' \href{http://dx.doi.org/10.1088/1475-7516/2023/02/012}{{\em
  JCAP} {\bfseries 02} (2023) 012},
  \href{http://arxiv.org/abs/2205.13227}{{\ttfamily arXiv:2205.13227
  [astro-ph.CO]}}.

\bibitem{GW}
M.~Mirbabayi, L.~Senatore, E.~Silverstein, and M.~Zaldarriaga, ``{Gravitational
  Waves and the Scale of Inflation},''
  \href{http://dx.doi.org/10.1103/PhysRevD.91.063518}{{\em Phys. Rev. D}
  {\bfseries 91} (2015) 063518},
  \href{http://arxiv.org/abs/1412.0665}{{\ttfamily arXiv:1412.0665 [hep-th]}}.

\bibitem{Qiu}
Y.~Qiu and L.~Sorbo, ``{Spectrum of tensor perturbations in warm inflation},''
  \href{http://dx.doi.org/10.1103/PhysRevD.104.083542}{{\em Phys. Rev. D}
  {\bfseries 104} no.~8, (2021) 083542},
  \href{http://arxiv.org/abs/2107.09754}{{\ttfamily arXiv:2107.09754
  [astro-ph.CO]}}.

\bibitem{Villadoro}
A.~Notari, F.~Rompineve, and G.~Villadoro, ``{Improved hot dark matter bound on
  the QCD axion},'' \href{http://arxiv.org/abs/2211.03799}{{\ttfamily
  arXiv:2211.03799 [hep-ph]}}.

\end{thebibliography}\endgroup
\end{document}